\documentclass[journal=ancac3,manuscript=article]{article}

\usepackage[version=3]{mhchem} 
\usepackage{epstopdf}
\usepackage{color,soul} 
\sethlcolor{yellow}
\usepackage{graphicx}
\usepackage{pdfpages}
\usepackage{setspace}
\usepackage[super,sort&compress,comma, numbers]{natbib}
\usepackage{caption}
\makeatletter
\renewcommand\@biblabel[1]{#1.}
\makeatother
\title{Integrated nanophotonics based on wire plasmons and atomically-thin material}
\author{Kenneth M. Goodfellow$^{1}$, Ryan Beams$^1$, Chitraleema Chakraborty$^2$, \\Lukas Novotny$^3$, \& A. Nick Vamivakas$^{1}$}

\date{}

\begin{document}

\maketitle

\begin{centering}

$^1$ Institute of Optics, University of Rochester, Rochester, New York 14627, USA

$^2$Materials Science,University of Rochester, Rochester, New York 14627, USA

$^3$ETH Z\"{u}rich, Photonics Laboratory, 8093 Z\"{u}rich, Switzerland

\end{centering}




\begin{abstract}

Photonic  integrated circuits are an enabling technology in modern communications systems. The continually increasing demands for higher-speed and lower operating power devices have resulted in the continued impetus to shrink photonic components.  In this work, we demonstrate a primitive nanophotonic integrated circuit element composed of a single silver nanowire and single-layer molybdenum disulfide (MoS$_2$) flake.  We show that nanowire plasmons can excite MoS$_2$ photoluminescence via direct plasmon-to-exciton conversion along the wire and plasmon-to-photon-to-exciton conversion at the MoS$_2$-covered wire end.  We also find that the reverse process is possible: MoS$_2$ excitons can decay into nanowire plasmons that can then be routed via the nanowire on-chip.  Finally, we demonstrate that the nanowire may serve the dual purpose of both exciting MoS$_2$ photoluminescence via plasmons and recollecting the decaying excitons.

\end{abstract}

\doublespace
As silicon photonic integrated circuits have continued to mature\cite{Rattner2010}, novel nanophotonic devices and nanomaterials are being explored for their potential in next-generation on-chip optical processing\cite{Sorger2012,Huang2014}. Particularly exciting is the possibility to engineer nanophotonic devices that both enhance light-matter interaction and support confined electromagnetic modes that can propagate in deeply subwavelength regions. Surface plasmon polaritons (SPPs)\cite{Novotny06a,maier07}, electromagnetic excitations that propagate along the interface between a metal and a dielectric, are a natural candidate for both integrated subwavelength light guiding and pronounced light-matter coupling\cite{BarnesNature2003,OzbayScience2006,MaierJAP2005,BozhevolnyiNature2006, OultonNatPhoton2008,PacificiNatPhoton2007,BeamsNano2013}.  An exemplary system in this regard are silver (Ag) nanowires\cite{DitlbacherPRL2005, LarocheAPL2006,  ZouAPL2010, WeiPNAS2013, ChangPRL2006, ManjavacaNano2009, SandersNano2006,  WangNano2011, LiNano2009, LiNano2010,FangNano2010} and, to date, the optical properties of individual Ag nanowires have been extensively studied\cite{SandersNano2006,LiNano2009,FangNano2010,LiNano2010,WangNano2011}. A step towards efficient  and compact nanophotonic circuitry is the integration of plasmonic waveguides that couple directly to on-chip sources, detectors, and modulators\cite{Huang2014}.  Initial steps have been made in coupling Ag nanowires with other nanostructures, such as quantum dots\cite{FedutikPRL2007, WeiNanoLett2009, GruberNano2013, AkimovNature2007}, fluorescent molecules\cite{ShegaiAPL2010}, and nitrogen-vacancy centers\cite{HuckPRL2011, KolesovNatPhys2009}. Furthermore, near-field coupling between these nanostructures and the wire allows for plasmons to be generated anywhere along the wire, not just at the ends\cite{WeiNanoLett2009, AkimovNature2007, ShegaiAPL2010, HuckPRL2011, KolesovNatPhys2009}.  Recently, an on-chip germanium field-effect transistor has exploited this near-field coupling to directly measure Ag nanowire plasmons\cite{FalkNP2009}.  

 Although there has been some investigation into graphene-nanowire hybrids for nanophotonic circuitry\cite{KimNano2012, QianACSNano2014}, the vast potential for two-dimensional atomically-thin materials in this realm is largely unexplored.   Single-layer molybdenum disulfide (MoS$_2$)\cite{MakPRL2010}, a semiconductor being explored for its photoluminescence\cite{SplendianiNano2010}, valley-selective properties\cite{ZengNatNano2012,MakNatNano2012,CaoNatCommun2012}, and potential as a transistor\cite{RadisavljevicNatNano2011} and photodetector\cite{YinACSNano2012,LopezSanchezNatNano2013}, is an ideal choice to couple with nanoplasmonic circuitry.  In this paper, we explore the nanophotonics of a MoS$_2$/Ag nanowire hybrid structure.  We demonstrate coupling between a single-layer MoS$_2$ flake and a single Ag nanowire.  We show that a plasmon excited at the uncovered end of the nanowire can propagate and excite MoS$_2$ photoluminescence (PL), both by direct plasmon-to-exciton conversion along the wire and by absorbing photons rescattering from the end of the wire.  We also demonstrate MoS$_2$ excitons can decay to generate Ag-nanowire plasmons. Finally, we show it is possible for the Ag nanowire to serve a dual role as both a channel for MoS$_2$ excitation and subsequent extraction of the decaying MoS$_2$ excitons. 

\begin{figure}
\centering
\includegraphics[width=160mm]{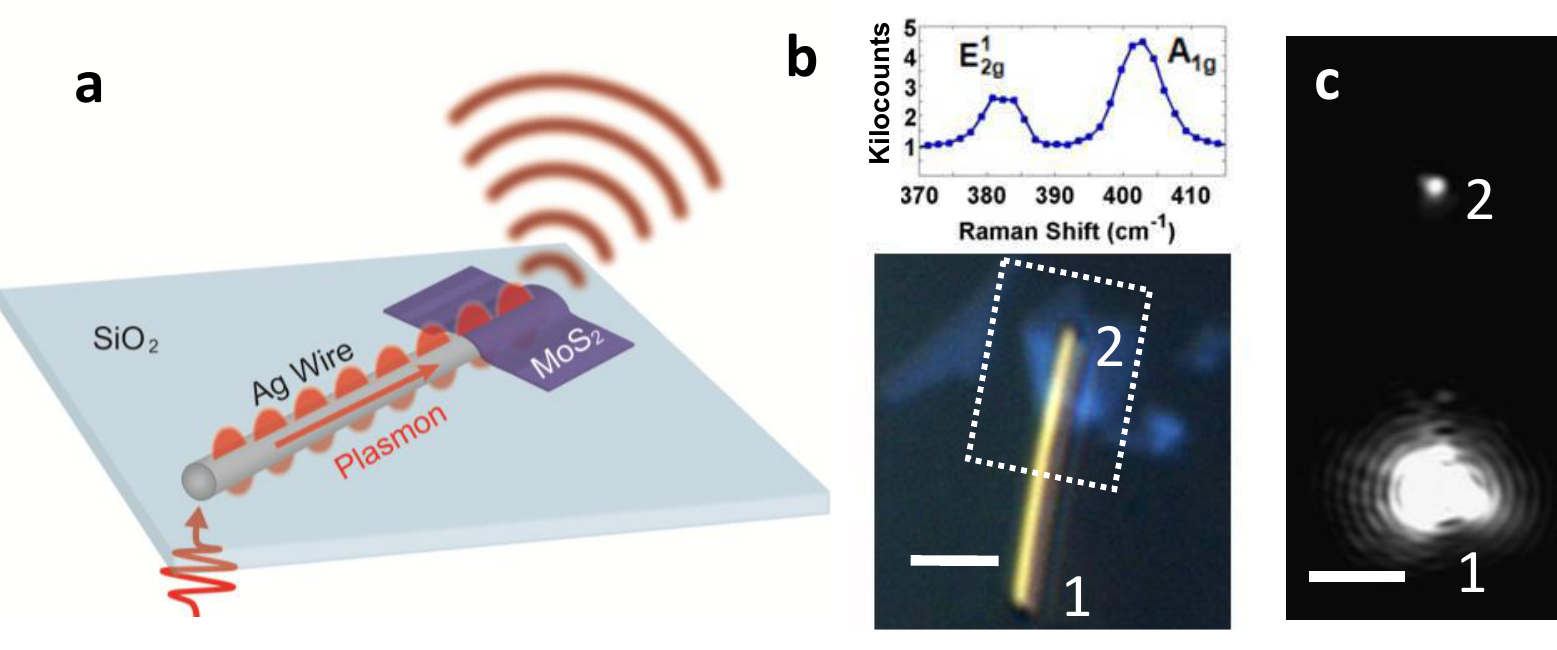}
\caption{$\textbf{Single-layer MoS$_2$/nanowire device}$. ($\textbf{a}$) Schematic of MoS$_2$/nanowire structure. ($\textbf{b}$) Top panel: Raman spectrum collected at the end of the wire in MoS$_2$.  $\lambda$ = 532 nm, power = 70 $\mu$W, integration time = 150 s. Bottom panel: Single-layer MoS$_2$ on a silver nanowire on glass after transfer of the flake. ($\textbf{c}$)  Demonstration of plasmon propagation along the Ag nanowire.  Light polarized parallel to the wire (bright spot) excites Ag nanowire plasmons that propagate along the wire and rescatter to the far field as photons (dimmer spot).  $\lambda$ = 635 nm, power = 20 $\mu$W.  All scale bars are 2 $\mu$m.
\label{Schematic}}
\end{figure}

\section*{Results}

\subsection*{MoS$_2$/nanowire hybrid device}Figure~\ref{Schematic}a presents an illustration of the fabricated MoS$_2$/Ag nanowire device.  An incident photon is converted to a  plasmon that propagates along the wire.  When the plasmon arrives at the MoS$_2$, the plasmon may either be converted to an exciton, resulting in frequency-shifted photon emission from the MoS$_2$, or it can be converted back to a photon at the end of the wire.  An optical micrograph of the hybrid device studied in this work is shown in the bottom panel of Fig.~\ref{Schematic}b.  The top panel is a Raman spectrum acquired at the overlap region between the end of the nanowire and the MoS$_2$.  The measured Raman spectrum reveals that the flake is single-layer MoS$_2$\cite{LeeACSNano2010}. See the Methods section for details on the fabrication of the device.

The charge-coupled device (CCD) image in Fig.~\ref{Schematic}c demonstrates plasmon propagation and photon re-emission.  Laser radiation ($\lambda$ = 635 nm), polarized parallel to the wire axis, is coupled from the far-field into the nanowire at the end labeled ``1'' in Fig.~\ref{Schematic}b using a 100$\times$ oil-immersion objective with numerical aperture (NA) of 1.3.  The power at the sample is 20 $\mu$W.  To reduce scattering and eliminate leakage radiation, the sample was covered in index-matching ($\textit{n}$ = 1.515) oil.   In order to convert a photon into an SPP, the laser must be focused onto one of the ends of the wire; this accounts for the momentum mismatch between the incoming photon and the plasmon\cite{SandersNano2006}.  Due to confinement of the SPP modes, smaller-diameter wires yield shorter 1/$e$ propagation lengths\cite{FalkNP2009, TakaharaOL1997}.  In addition, the SPP 1/$e$ propagation length increases as the optical excitation wavelength increases\cite{BarnesNature2003}.  The wires used in our study support two lower-order modes.  When the incident light is polarized parallel to the wire, the light couples to the lowest-order, $m=0$ SPP mode (Fig.~\ref{Schematic}c), and light oriented perpendicular to the wire couples to the $m = 1$ mode\cite{LiNano2010} (see Supplementary Fig.~1e and 1f).  The in-coupling efficiency is always greater for the $m = 0$ mode than for the $m = 1$ mode, but the 1/$e$ propagation length of the $m = 1$ mode becomes longer for larger diameter wires, explaining why SPP propagation still occurs when incoming light is polarized perpendicular to the wire.  For a wire of this length, we calculate the efficiency of photon re-emission at the end of the wire after plasmon propagation to be around 0.008\% to 0.012\%. See Supplementary Discussion 1 and Supplementary Fig.~1 for a discussion of photon re-emission efficiencies.

\begin{figure}
\centering
\includegraphics[width=160mm]{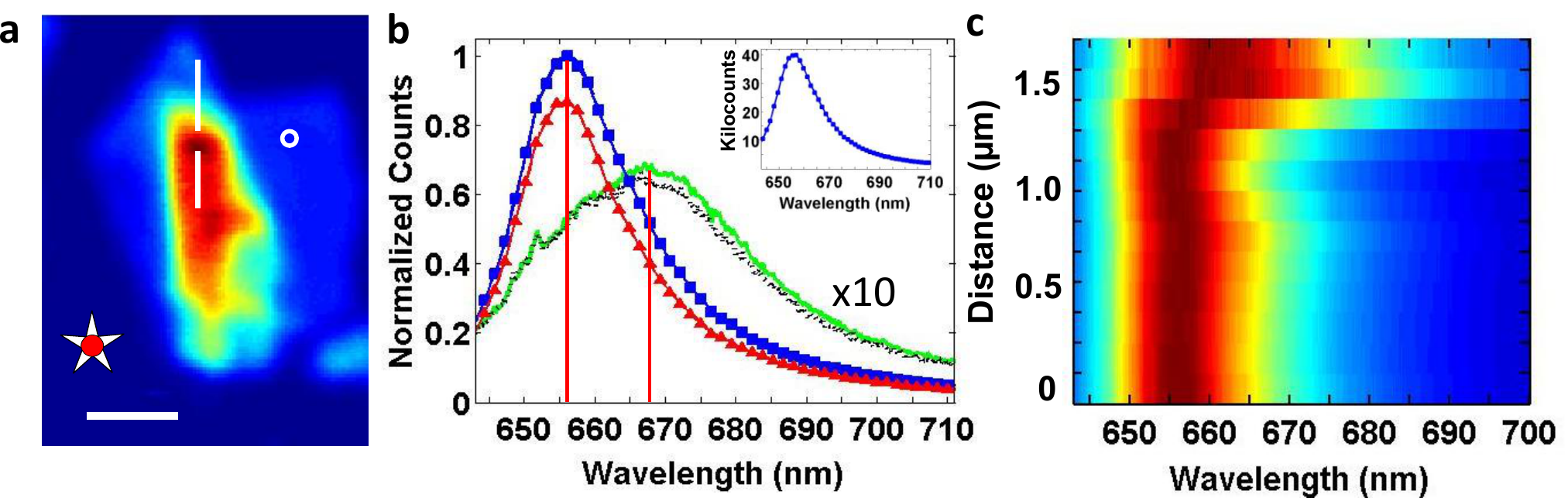}
\caption{$\textbf{Photoluminescence from the MoS$_2$/wire device}$. ($\textbf{a}$) Confocal fluorescence image of the MoS$_2$/wire device from the dotted white box in Fig.~\ref{Schematic}b. The red dot and white star represent no displacement between the excitation and collection focal volume, respectively.  Scale bar is 1 $\mu$m. ($\textbf{b}$) Normalized photoluminescence spectra for the end of the wire covered by MoS$_2$ excited with light polarized parallel (blue decorated with squares) and perpendicular (red decorated with triangles) to the wire. Spectra of the bare wire were also acquired using parallel (green) and perpendicular (black) excitation polarizations. The spectra for the MoS$_2$ on substrate are enhanced by a factor of 10 for clarity.  Inset: Absolute photoluminescence counts for the end of the wire in MoS$_2$ with light polarized parallel to the wire.  ($\textbf{c}$) Spectra taken in ~150 nm steps along the line in (a).  For all images, $\lambda$ = 633 nm, power = 5 $\mu$W.  For all spectra, integration time = 40 s.
\label{SCAN}}
\end{figure}

\subsection*{Optical interaction between the nanowire and MoS$_2$}Coupling between far-field photons, SPPs, and the single-layer MoS$_2$ was studied using an inverted confocal microscope utilizing an oil-immersion objective (NA = 1.4). See the Methods section for more details. A fluorescence image of the single-layer flake on the wire from Fig.~\ref{Schematic}b (region is outlined with dashed white box) is shown in Fig.~\ref{SCAN}a. For photon counting images, we adopt the convention of using a solid red circle to indicate the excitation and a white star to represent the approximate center of the collection focal volume, respectively.  For this data set, the excitation and collection focal volumes are coincident so the dot and star overlap indicating no displacement between excitation and collection; however, later images introduce displacement of the excitation with respect to the collection.  In  Fig.~\ref{SCAN}a, we observe strong direct band gap PL from the MoS$_2$ flake, characteristic of single-layer, as well as a large increase in counts in the area where the flake overlaps with the wire. To investigate the origin of the increased counts, Fig.~\ref{SCAN}b presents spectra acquired with the excitation polarization parallel and perpendicular to the wire on the bare flake (white circle in Fig.~\ref{SCAN}a) and at the wire-flake overlap region (break in the solid vertical line in Fig.~\ref{SCAN}a).  Spectra taken at the wire-flake overlap region (bare flake) with excitation polarized parallel to the wire is shown in blue (green), and the red (black) curve shows the perpendicular polarization case.  There can be a number of contributing effects to this enhancement.  First, by removing direct contact with the substrate, MoS$_2$ fluorescence is known to increase\cite{MakPRL2010,ScheuschnerPRB2014}.  Second, the enhancement of the MoS$_2$ fluorescence  when the excitation is polarized parallel to the nanowire is a manifestation of an antenna-like enhancement of the excitation field. In contrast, the PL intensity from the MoS$_2$ on the substrate does not exhibit dependence on the excitation polarization.  The enhancement in MoS$_2$ fluorescence in the vicinity of the region where the Ag nanowire end overlaps with the MoS$_2$  flake is consistent on all devices we have fabricated (see Supplementary Fig.~2 for data from four other single-layer devices, Supplementary Fig.~3 for a bilayer sample, and Supplementary Fig.~4 for a device that exhibits a 40-fold enhancement).  

In addition to the enhancement, there is a clear spectral shift in the peak of the PL for the MoS$_2$ over the wire (656 nm) compared to on the substrate (668 nm).  We attribute this shift to the MoS$_2$ flake not being in direct contact with the Ag nanowire as a result of the transfer process.  The main PL peak of single-layer MoS$_2$ consists of two peaks: the A peak centered at 655 nm attributed to uncharged excitons and the A$^-$ peak centered around 670 nm due to negatively-charged trions\cite{MakNatMater2012}.  It has been reported that interaction of the MoS$_2$ with the substrate suppresses exciton emission due to doping\cite{ScheuschnerPRB2014}.  When removed from the substrate, the A peak becomes dominant.  This shift is not due to strain, as strain would redshift the spectra\cite{ConleyNano2013}. For comparison, Fig.~\ref{SCAN}c presents  MoS$_2$ spectra as we measure along the line in Fig.~\ref{SCAN}a, starting from the bottom.  Each spectrum is independently normalized.  The spectral position of the peak is consistent along the wire covered by the MoS$_2$, and as the collection region moves off of the wire, the peak redshifts.  See Supplementary Fig.~5 for spatially-resolved Raman spectra of the same line cut and both Raman and PL along the line orthogonal to the vertical white line in Fig.~\ref{SCAN}a. 

\begin{figure}
\centering
\includegraphics[width=160mm]{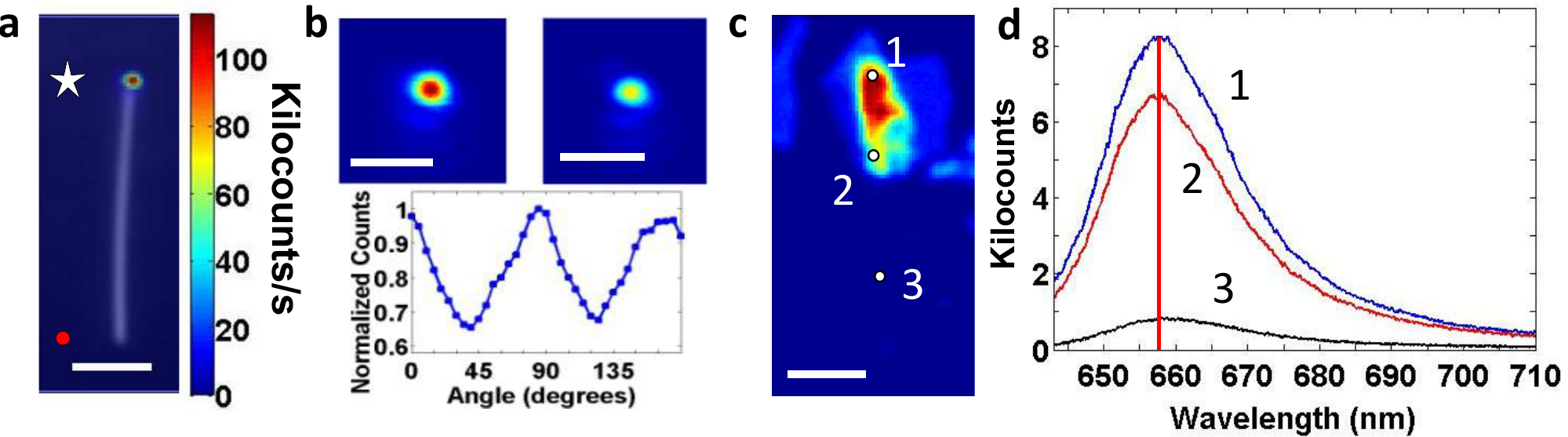}
\caption{$\textbf{Plasmonic excitation of single-layer MoS$_2$}$. ($\textbf{a}$) Fluorescence image resulting from displacing the collection and excitation focal volumes by the length of the wire. The displacement between the red dot (excitation) and white star (collection) indicates this distance.  A CCD image of the device is overlayed with the fluorescence image.  Scale bar is 2 $\mu$m.  ($\textbf{b}$) Top left (right) panel: Feature in (a) when light is polarized parallel (perpendicular) to the wire.  Scale bar in each is 1 $\mu$m.  Bottom panel: Normalized polarization contrast of the MoS$_2$ fluorescence as a function of excitation polarization angle with respect to the nanowire axis.  0$^{\circ}$ corresponds to polarization parallel to the wire.  ($\textbf{c}$) Confocal fluorescence image of the sample constructed with no displacement of the laser excitation.  The numbered dots indicate the locations of spectra in ($\textbf{d}$) when the laser excitation is located at the uncovered end of the wire.  Scale bar is 2 $\mu$m.  For all images, $\lambda$ = 633 nm, power = 5 $\mu$W.  For all spectra, integration time = 40 s.
\label{Misaligned}}
\end{figure}

\subsection*{Plasmonic excitation of MoS$_2$}To explore plasmon excitation of MoS$_2$ PL, the collection and excitation focal volumes are displaced vertically by the length of the wire.   Figure~\ref{Misaligned}a shows the resulting fluorescence image when the sample is scanned in this configuration with the laser polarized parallel to the wire.   A CCD image of the MoS$_2$/wire structure is overlayed on this image. The prominent feature in the fluorescence image results when the laser excites plasmons at the end of the wire opposite the MoS$_2$, which propagate along the wire and excite MoS$_2$ PL.  Figure~\ref{Misaligned}b displays scans of this feature when light is polarized parallel (upper left panel) and perpendicular (upper right panel) to the wire.  There is a reduction in intensity when the light is polarized perpendicular to the wire, suggesting that this feature is the result of plasmon propagation. The bottom panel of Fig.~\ref{Misaligned}b shows the normalized polarization dependence of the signal with a visibility of 21\%. While the large-diameter Ag wire has modes that can be excited with an incident field polarized parallel or perpendicular to the wire, the observed modulation indicates that the coupling is stronger when the excitation is parallel.  The largest visibility that we observed on a device was 80\% (see Supplementary Fig.~6). Simulations using the finite-difference time-domain (FDTD) method in Lumerical to further investigate the plasmon excitation are shown in Supplementary Fig.~7.

We anticipate that plasmon-excited MoS$_2$ PL is not limited to the end of the wire.  To investigate this, the displacement between the laser excitation and collection was adjusted to be a fraction of the wire length.  The sample is then translated so that the laser excitation is at the uncovered end of the wire.  For reference, Fig.~\ref{Misaligned}c shows a fluorescence scan of the full sample with circles to mark the effective positions of the spectral collection along the wire.  Figure~\ref{Misaligned}d presents the spectra corresponding to each of these points, starting from the top circle, labeled ``1'', and walking downward in Fig.~\ref{Misaligned}c.  We observe that the PL is strongest near the end of the wire.  However, we also obtain significant signal over the entire length that the wire is covered by the MoS$_2$.  This is suggestive of two mechanisms at play.  First, plasmons that propagate to the end of the wire are rescattered as photons and reabsorbed by the MoS$_2$, exciting an exciton.  The electron-hole recombination then produces the PL signal.  Second, plasmons in the wire are directly converted to excitons in the MoS$_2$, which then fluoresces.  The exact separation between the nanowire and the MoS$_2$ flake determines the efficiency of direct plasmon absorption.

\begin{figure}
\centering
\includegraphics[width=160mm]{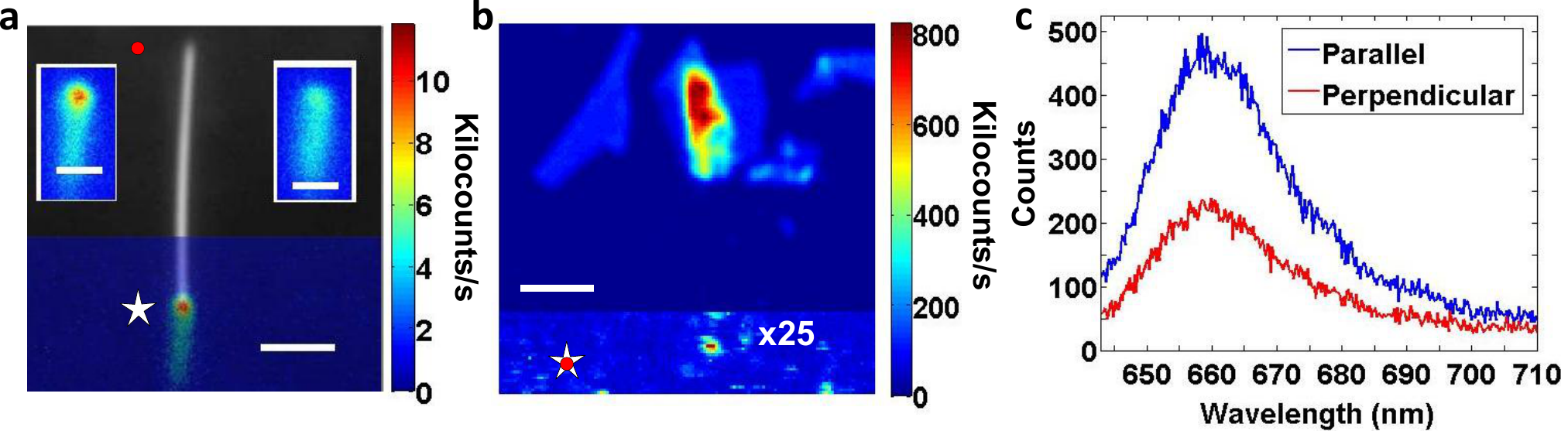}
\caption{$\textbf{Plasmon extraction of single-layer MoS$_2$ fluorescence}$. ($\textbf{a}$) Image resulting from displacing the collection and excitation focal volumes by the length of the wire.  The displacement between the red dot (excitation) and white star (collection) displays this distance.  Opposed to Fig.~\ref{Misaligned}, in this case the excitation is located at the MoS$_2$ end, and the collection is at the uncovered wire end.   The image is overlayed with an image of the structure.  Scale bar is 2 $\mu$m.  Left (right) inset: Feature when light is polarized parallel (perpendicular) to the wire.  Scale bar in each is 1 $\mu$m. $\lambda$ = 635 nm, power=20 $\mu$W. ($\textbf{b}$) Confocal fluorescence image of the sample with no displacement in the excitation and with the uncovered end of the wire rescaled.  Scale bar is 2 $\mu$m.  ($\textbf{c}$) Spectra collected at the rescaled feature in (a) for light polarized parallel (blue) and perpendicular (red) to the wire.  For (b) and (c), $\lambda$ = 633 nm, power = 5 $\mu$W, spectral integration time = 40 s.  
\label{MoS2exciton}}
\end{figure}

\subsection*{Plasmon extraction from MoS$_2$ fluorescence}In addition to plasmons exciting the MoS$_2$, the reverse process can also occur; that is, excitons in the MoS$_2$ can be converted to plasmons that propagate along the wire and are rescattered as photons. To demonstrate that MoS$_2$ fluorescence can couple to Ag nanowire plasmons, the excitation is aligned with the overlap region of the MoS$_2$ flake/nanowire end, and the collection focal volume is aligned to the uncovered nanowire end (the reverse configuration of Fig.~\ref{Misaligned}).  Figure~\ref{MoS2exciton}a shows the resulting fluorescence image when the sample is scanned in this configuration with the excitation laser polarized parallel to the wire.  Again, a CCD image of the MoS$_2$/wire structure is overlayed on this image.  Compared to the localized feature in Fig.~\ref{Misaligned}a, the present image shows an attribute that extends beyond the end of the wire.  This is suggestive of plasmonic excitation along the MoS$_2$/wire interface, not just at the covered end.  As the laser excitation scans over the  MoS$_2$/wire interface, photons that re-emerge from the uncovered end are still detected by the confocal volume of the APD.  If plasmons could only be excited at the end of the wire and not along the length of the MoS$_2$/wire overlap, the attribute in this image would look similar to the feature in Fig.~\ref{Misaligned}a.   The insets display scans of the feature when the excitation light is polarized both parallel (upper left) and perpendicular (upper right) to the wire.  Evident from comparing the two insets is an increase in the emission from the uncovered end when the excitation is parallel to the nanowire axis.  The enhancement again suggests the nanowire provides an antenna-like enhancement of the excitation. Because MoS$_2$ absorption does not prefer a linear polarization, any mismatch results from excitation effects.

Finally, we demonstrate it is possible to use the Ag nanowire both as a channel for near-field excitation of the MoS$_2$ flake and to recollect the resultant MoS$_2$ fluorescence.  The re-excited plasmons, at the MoS$_2$ photon energy, can propagate back along the wire and rescatter to the far-field as photons.  Figure~\ref{MoS2exciton}b shows a fluorescence image of the entire sample with the excitation and collection aligned. For this image, we have rescaled the end of the wire not covered by the MoS$_2$.  There is a pronounced fluorescence feature at the excitation end of the wire that is stronger than the background.  To investigate this feature, spectra were collected for light polarized parallel and perpendicular to the wire (Fig.~\ref{MoS2exciton}c). The spectra reveal this is indeed PL from the MoS$_2$ flake; see Fig.~\ref{SCAN}b for comparison.  The pronounced polarization contrast in the two different excitation directions suggests that the Ag nanowire plasmons mediate this excitation and collection process. 

In summary, we have demonstrated photonic and plasmonic interactions between an individual Ag nanowire and single-layer MoS$_2$.  We found it is possible to excite MoS$_2$ with Ag nanowire plasmons as well as convert decaying MoS$_2$ excitons into Ag wire plasmons.  This first step shows that there is pronounced nanoscale light-matter interaction between plasmons and atomically-thin material that can be exploited for nanophotonic integrated circuits. A natural next step is the creation of a near-field detector based on MoS$_2$ as well as MoS$_2$ light-emitting diodes coupled to on-chip nanoplasmonic circuitry.

\section*{Methods}

\subsection*{Sample fabrication}
A solution of silver nanowires (Nanostructured and Amorphous Materials, Inc.) with 386 nm average diameter and 8.6 $\mu$m average length is diluted at a 2000:1 ratio of ethanol to solution, and a small amount ($\sim$150 $\mu$L)  is stamped onto a clean silica coverslip using a polycarbonate membrane filter (5 $\mu$m pore diameter, SPI Supplies, Inc.).   The coverslip was placed in a petri dish with a hole bored in it and held in place with rubber cement for the transfer. On a separate silicon substrate with 270 nm of oxide (Si/SiO$_2$), we exfoliated MoS$_2$ from the bulk (SPI Supplies, Inc.) by micromechanical cleavage using adhesive tape (Semicorp).  The optical interference due to the thickness of the oxide allows for identification of single-layers\cite{BenameurNano2011}, which is also confirmed by Raman spectroscopy (Fig.~\ref{Schematic}b top panel)\cite{LeeACSNano2010}.  The flakes were lifted off using a poly(methyl methacrylate) (PMMA)-based technique\cite{ReinaJPCC2008, SundaramNano2013}. After a candidate MoS$_2$ flake is found on the Si/SiO$_2$, two coats of 495K PMMA and one coat of 950K PMMA were spun onto the Si/SiO$_2$; the sample was baked for 5 minutes at 105$^{\circ}$ C then immersed in 1 M NaOH at 80$^{\circ}$ C until release begins (about 10 minutes). It was transferred to deionized (DI) water, where the PMMA film was detached with assistance from tweezers.  The petri dish with the wire sample was then filled with DI water, and a suitable wire was found using an inverted microscope with a long working distance objective.  The PMMA film was transferred to the petri dish, and a post with a teflon-coated end attached to a micropositioner was brought into contact with the PMMA.  Using a Harvard PhD 2000, the water was pumped from the petri dish; the objective focus could be adjusted to position the flake over the wire.  The sample dried before immersion in an acetone bath to dissolve the PMMA.  

\subsection*{Optical characterization} 
The samples were characterized with an inverted microscope equipped with an oil-immersion objective.  A nanopositioning stage (Mad City Labs, Inc.) was used to scan and position the sample.  The sample was characterized using a 532 nm wavelength laser for Raman spectroscopy or a 633 nm wavelength laser for photoluminescence and plasmon propagation measurements.  Excitation polarization was controlled by a half-wave plate.  The signal from the sample was sent to either an APD or a spectrometer.  Longpass filters to block the laser line were used in front of both detectors.  For some of the fluorescence images, a similar second set-up with a 635 nm wavelength laser was used.  Laser power of 5 $\mu$W was used in Fig.~\ref{SCAN}, Fig.~\ref{Misaligned}, and Fig.~\ref{MoS2exciton}b and \ref{MoS2exciton}c.  Laser power of 20 $\mu$W was used in Fig.~\ref{Schematic}c and Fig.~\ref{Misaligned}a. Laser power of 70 $\mu$W was used in Fig.~\ref{Schematic}b.

%

\subsection*{Acknowledgements} The authors acknowledge support from the Institute of Optics, the U. S. Department of Energy (grant DE-FG02-05ER46207), the National Science Foundation IGERT program,  and the National
Science Foundation (DMR-1309734).

\subsection*{Author Contributions} R. B., L. N., and A. N. V. conceived the research.  K. G. and C. C. fabricated the samples.  K. G. and R. B. conducted the measurements.  C. C. conducted the simulations.  All authors discussed the data and wrote the manuscript. 

\subsection*{Competing Financial Interests} The authors declare no competing financial interests.



\section*{Supplementary material}

# Supporting information for:

# Integrated nanophotonics based on wire plasmons

# and atomically-thin material

Kenneth M. Goodfellow[1], Ryan Beams[1], Chitraleema Chakraborty[2], Lukas Novotny[3], and A. Nick Vamivakas[1]

[1]Institute of Optics, University of Rochester, Rochester, New York 14627, USA
[2]Materials Science, University of Rochester, Rochester, New York 14627, USA
[3]Photonics Laboratory, ETH Zürich, 8093 Zürich, Switzerland

## Supplementary Discussion 1:

## Estimation of photon re-emission efficiency from wires

Supplementary Figure 1a displays a bare wire of length 5 µm (wire in text was about 7 µm long). Supplementary Fig. 1b shows a scanning electron microscope (SEM) image of a different wire to show the tapered end geometry. We show plasmon propagation resulting in photon reemission for the wire in Supplementary Fig. 1a in Supplementary Fig. 1c with the laser polarized parallel to the wire, and Supplementary Fig. 1d displays the image resulting from displacing the collection and excitation focal volumes by a transverse length equal to the wire length. These images are contrasted to Supplementary Fig. 1e and 1f, in which the excitation is perpendicular to the wire. The filters in front of the APD are an optical density filter to cut three orders of magnitude of light (OD3) and a 633/10 bandpass. From this, we can estimate the photon-to-photon conversion efficiency. For this wavelength and power, the input photon flux is approximately $6.4 \times 10^{13}$ photons/s. At the APD, the photon flux is approximately $2.5 \times 10^{6}$ photons/s for the parallel excitation. Accounting for the OD3 filter (1000x reduction) and a factor of 4-6x for losses in the system (including a beamsplitter and a pinhole apparatus), we arrive at an efficiency of about 0.015% to 0.023% for this wire. The ratio of photon flux out to photon flux in takes the form of $R = Ae^{\frac{-L}{L_o}}$, where $A$ represents a factor encompassing the photon-to-plasmon and plasmon-to-photon coupling efficiencies, and $L_o$ is the propagation length. Using simulation data, we estimate $L_o$ to be about 3.1 µm for our wires. Using this value and the calculated efficiency for the wire in Supplementary Fig. 1a, we calculate $A$ to be between $7.37 \times 10^{-4}$ and $1.13 \times 10^{-3}$. Applying these values to the wire in the main text, we can estimate the efficiency to be about 0.008% to 0.012%.

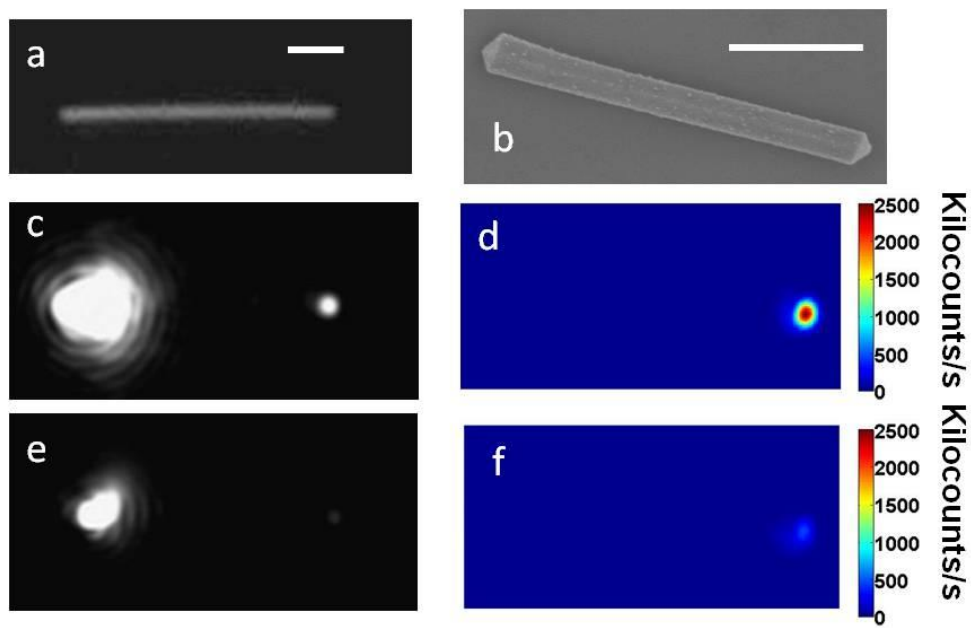

**Supplementary Figure 1. Plasmon propagation and photon reemission for a bare wire.** (**a**) CCD image of a bare wire. (**b**) SEM image of a bare wire. For (a) and (b), scale bars are 1 μm. (**c**) Demonstration of plasmon propagation along the nanowire in (a) and photon reemission at the opposite end. (**d**) Image resulting from displacing the collection and excitation focal volumes by a transverse length equal to the wire length. (**e**) and (**f**) The corresponding images to (b) and (c) for the polarization parallel to the wire. For (c) through (f), λ = 635 nm, Power = 20 μW.

# Additional MoS₂/wire devices

Supplementary Figure 2 shows other single-layer MoS$_2$/wire devices. Similar to the device presented in the text, we observed increased MoS$_2$ PL counts at the MoS$_2$-wire interface in four other devices.

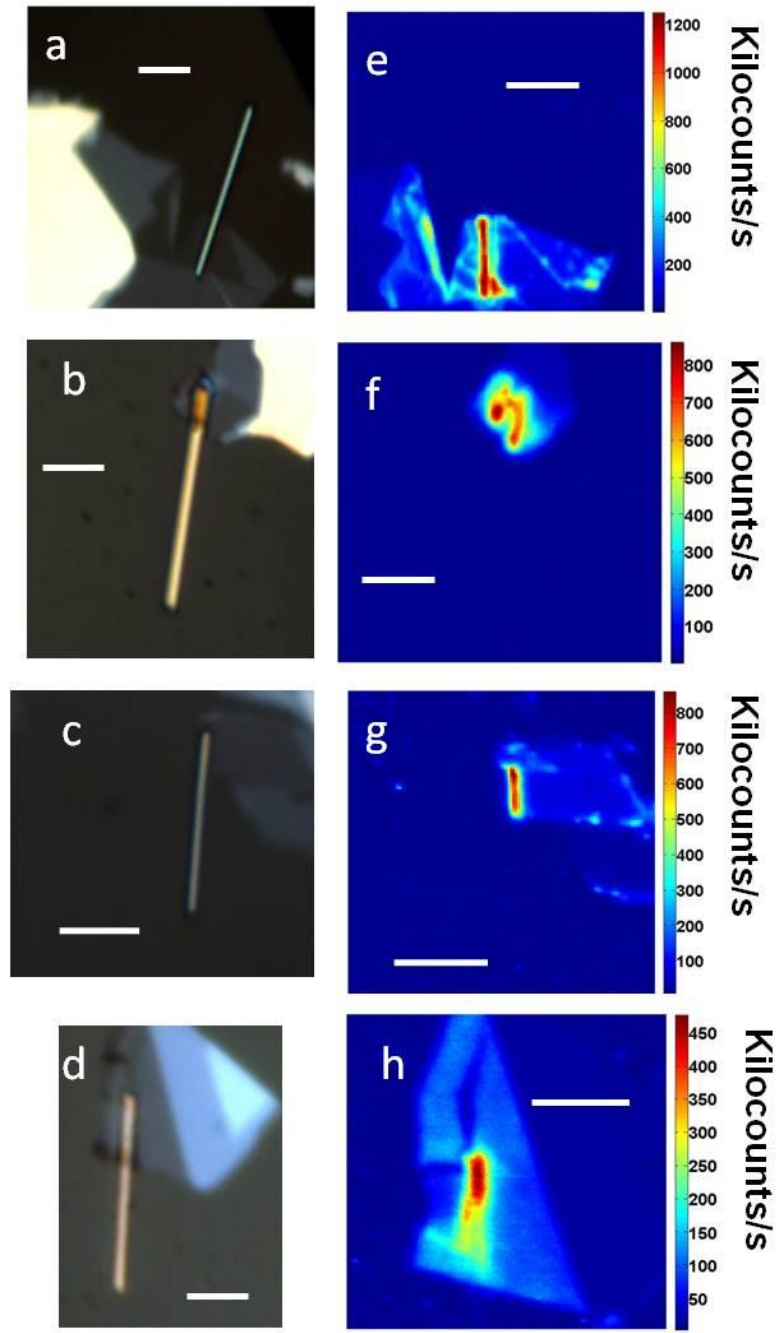

**Supplementary Figure 2. Additional MoS$_2$/wire devices.** (**a**)-(**d**) Wide-field images of other fabricated single-layer MoS$_2$/wire devices. (**e**)-(**h**) Corresponding fluorescence images of the devices shown. Scale bars for all images are 3 μm. For (e) and (f), λ =6 33 nm, Power = 5 μW. For (g) and (h), λ = 532 nm, Power = 5 μW.

# Bilayer MoS$_2$/wire device

Supplementary Figure 3 displays data for a bilayer MoS$_2$/wire device. Supplementary Fig. 3a show a wide-field image of this device, and a corresponding fluorescence image is shown in Supplementary Fig. 3b. Again, there is an enhancement in the region of MoS$_2$/wire overlap. Supplementary Fig. 3c displays spatially-resolved Raman spectra for the line across the sample in Supplementary Fig. 3b. The separation between the peaks confirms bilayer, and we see an enhancement in the signal over the wire/MoS$_2$. The right end of the line corresponds to the bottommost spectrum. The spectra are taken about 150 nm from each other. Supplementary Fig. 3d displays photoluminescence spectra for the end of the wire covered by MoS$_2$ for light polarized parallel (blue) and perpendicular (red) to the wire. These spectra are taken with no displacement in the focal volumes of the excitation and collection. We see a peak around 760 nm corresponding to the indirect band gap. Supplementary Fig. 3e shows the fluorescence image resulting when the collection and excitation focal volumes are displaced by a distance corresponding to the wire length. The image is overlayed on a CCD image of the device. The prominent feature is a result of exciting at the wire end opposite the MoS$_2$. The normalized polarization response as a function of angle rotated on the half-wave plate is shown in Supplementary Fig. 3f.

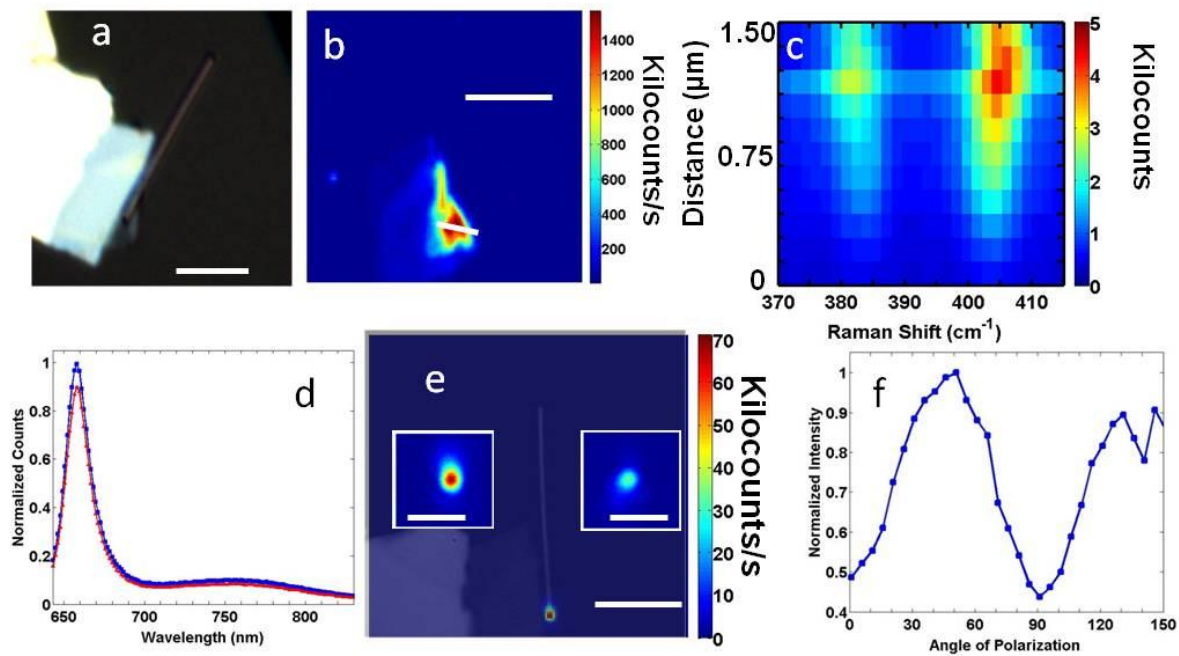

**Supplementary Figure 3**. **Bilayer MoS$_2$/wire device.** (**a**) Wide-field image for a bilayer MoS$_2$/wire device. Scale bar is 4 µm. (**b**) Fluorescence image of the device in (a). Scale bar is 4µm. (**c**) Raman spectra taken along the linecut in (b). For these spectra, λ = 532 nm, Power = 70 µW, integration time per spectrum = 150 seconds. (**d**) Normalized spectra taken at the MoS$_2$/wire for light polarized parallel (blue) and perpendicular (red) to the wire. (**e**) Fluorescence image resulting from displacing the collection and excitation focal volumes by a transverse length equal to the wire length. A CCD image of the device is overlayed with the fluorescence image. Scale bar is 4 µm. Left (right) inset: Feature when light is polarized parallel (perpendicular) to the wire. Scale bar in each is 1µm. For (d) and (e), λ = 633 nm, Power = 5 µW. (**f**) Normalized polarization contrast of the MoS$_2$ fluorescence as a function of excitation polarization angle. 0° corresponds to polarization perpendicular to the wire.

# Device with large enhancement

Supplementary Figure 4 displays data for the MoS$_2$/wire device that had the highest enhancement. This is the same device that was shown in Supplementary Fig. 2c and 2g. Supplementary Fig. 4a presents a wide-field image of the region of interest, and the corresponding fluorescence image is shown in Supplementary Fig. 4b. Spectra taken at the MoS$_2$/wire interface as well as on the MoS$_2$ on substrate ~2 µm from the wire are shown in Supplementary Fig. 4c. In addition to the spectral shift described in the main text, we also notice an increase of ~40x between the height of the two peaks.

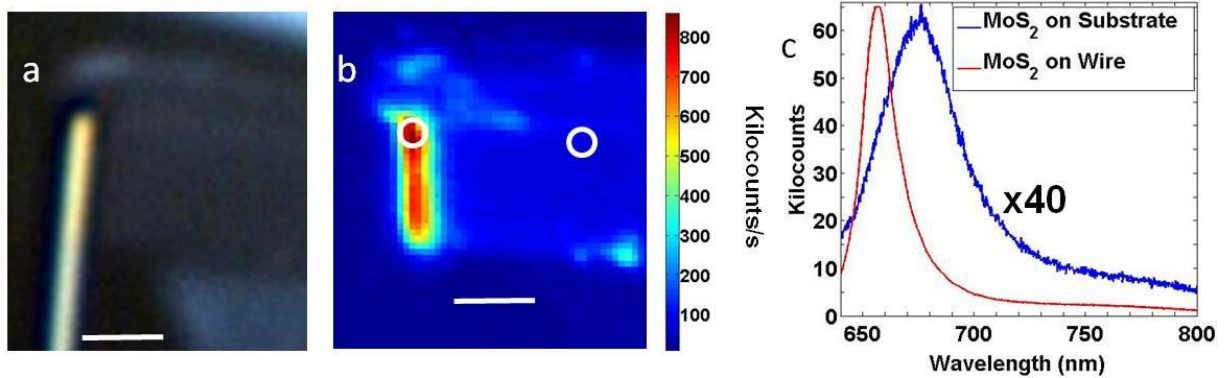

**Supplementary Figure 4. Device with large enhancement.** (**a**) Wide-field image of the single-layer MoS$_2$/wire device. (**b**). Fluorescence image of the device shown in (a). Scale bars for (a) and (b) are 1 µm. The two white circles correspond to the locations of the spectra in (**c**). For (b) and (c), λ = 532 nm, power = 5 µW. For (c), spectral integration time = 40 s.

# Spatially-resolved Raman spectra and photoluminescence for sample in the main text

Supplementary Fig. 5 shows spatially-resolved spectra of Raman measurements and photoluminescence. Supplementary Fig. 5a shows the same fluorescence image as shown in Figure 2a in the main text but with two linecuts at the end of the wire. Supplementary Fig. 5b shows photoluminescence spectra corresponding to the horizontal line. The left end of the line corresponds to the bottommost spectrum. Again, we see the spectral blueshift corresponding to the $MoS_2$/wire interface. Supplementary Fig. 5c and 5d display Raman spectra corresponding to the horizontal line and the vertical line, respectively. There is an enhancement in the signal for the $MoS_2$/wire overlap, but there is no shift in either peak.

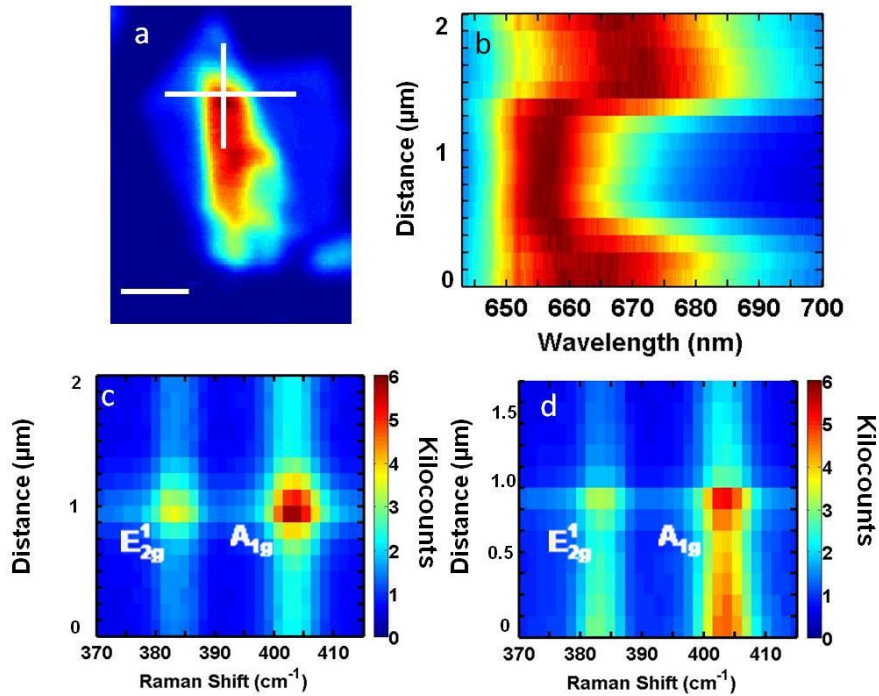

**Supplementary Figure 5. Spatially-resolved Raman spectra and photoluminescence for sample in the main text.** (**a**) Confocal fluorescence image of the MoS2/wire structure from Figure 2a in the main text. Scale bar is 1µm. (**b**) Photoluminescence spectra taken across the horizontal line in (a). For these spectra, λ = 633 nm, Power = 70 µW, integration time per spectrum = 40 s. (**c**) Raman spectra taken across the horizontal line in (a). (**d**) Raman spectra taken along the vertical line in (a). For (c) and (d), λ = 532 nm, Power = 70 µW, integration time per spectrum = 150 s.

# High-visibility single-layer MoS$_2$/wire device

Supplementary Figure 6a shows a wide-field image for another single-layer MoS$_2$/wire device (same device from Supplementary Fig. 2b), and the corresponding fluorescence image is shown in Supplementary Fig. 6b. Supplementary Fig. 6c shows the fluorescence image resulting when the collection and excitation focal volumes are displaced by a distance corresponding to the wire length. The image is overlayed on a CCD image of the device. The prominent feature is a result of exciting at the wire end opposite the MoS$_2$. The normalized polarization response as a function of angle rotated on the half-wave plate is shown in Supplementary Fig.6. We see that this sample has a much higher contrast between the two orthogonal polarizations and has a visibility of about 80%.

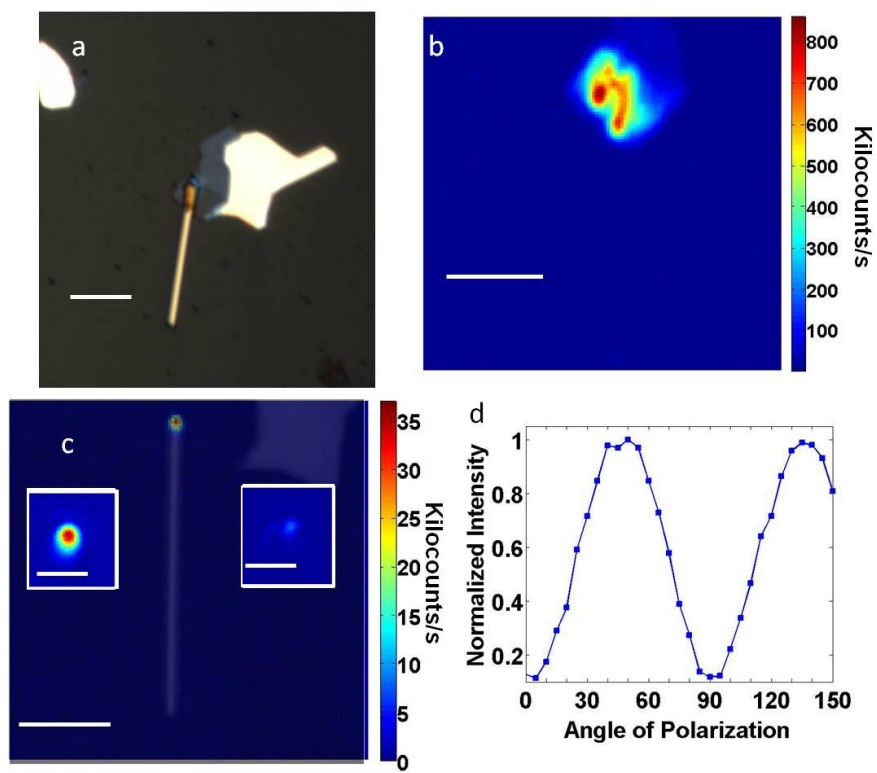

**Supplementary Figure 6. High-visibility single-layer MoS$_2$/wire device.** (**a**) Wide field image for another single-layer MoS$_2$/wire device. Scale bar is 4 μm. (**b**). Fluorescence image of the device in (a). Scale bar is 4μm. (**c**) Fluorescence image resulting from displacing the collection and excitation focal volumes by a transverse length equal to the wire length. A CCD image of the device is overlayed with the fluorescence image. Scale bar is 4μm. Left (right) inset: Feature when light is polarized parallel (perpendicular) to the wire. Scale bar in each is 1μm. For (b) and (c), λ = 633 nm, Power = 5 μW. (**d**) Normalized polarization contrast of the MoS$_2$ fluorescence as a function of excitation polarization angle. 0° corresponds to polarization perpendicular to the wire.

# FDTD simulation of the MoS$_2$/nanowire structure

Supplementary Figure 7 displays results of a simulation of the MoS$_2$/wire device. Supplementary Fig. 7a shows the simulation interface. A 380 nm silver nanowire is on a silica substrate with MoS$_2$ on one end of the wire. For the MoS$_2$, we assume a refractive index of 6.[1] Supplementary Fig. 7b shows the electric field profile at the beginning of the simulation with excitation at the uncovered end of the wire. The flake is outlined with a dotted line. The field profile after the simulation is shown in Supplementary Fig. 7c for the excitation end (labeled 1 in Supplementary Fig. 7a) and in Supplementary Fig. 7d for the MoS$_2$ end (labeled 2 in Supplementary Fig. 7a). We notice that the field in Supplementary Fig. 7d has non-zero components in the MoS$_2$ located directly at the wire as well as a short distance away. This is similar to what was observed in the experiments.

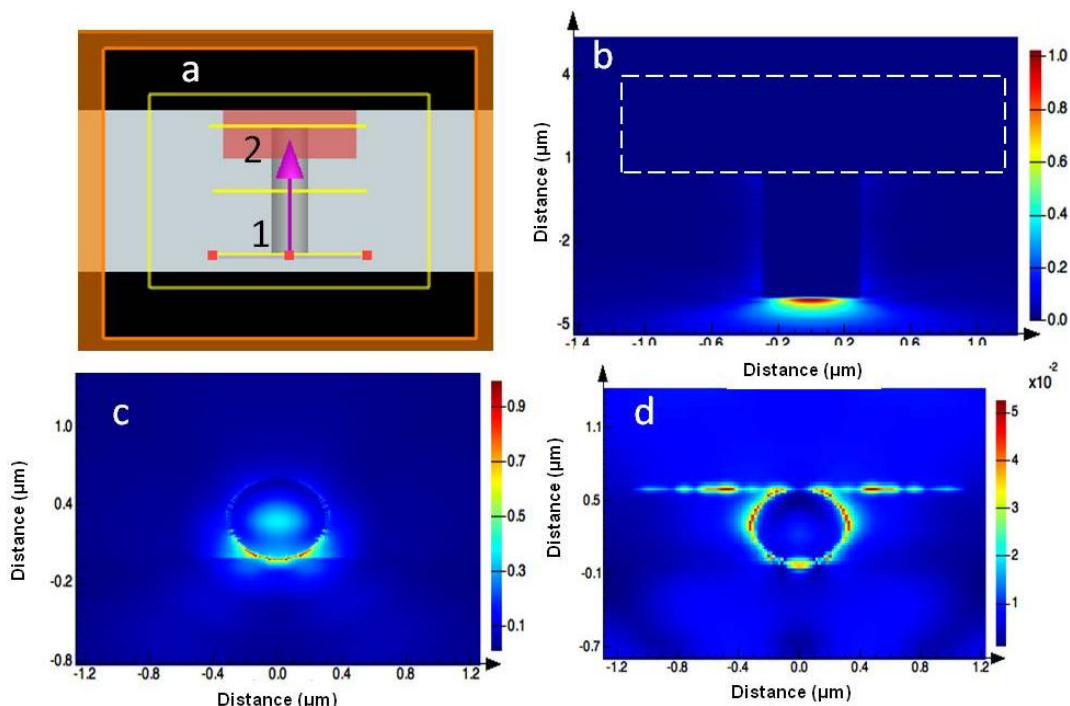

**Supplementary Figure 7. FDTD simulation of the MoS$_2$/nanowire structure.** (**a**) Interface for the FDTD simulation (Lumerical Solutions, Inc.), representing a 380 nm diameter, 8 μm long silver nanowire with MoS$_2$ on one end. The substrate is silica (n=1.5). (**b**) Top view of the electric field profile showing excitation at the uncovered end. (**c**) Cross-section view of electric field profile at the position marked "1" in (a). (**d**) Cross-section view of electric field profile at the position marked "2" in (a).

# Supplementary Reference



\end{document}